\documentclass[aps,pra,reprint,superscriptaddress]{revtex4-2}
\usepackage{amsmath,amssymb}
\usepackage{amsfonts}
\usepackage{subfigure}
\usepackage{bm}
\usepackage{graphicx}
\usepackage{appendix}
\usepackage{multirow}
\usepackage{textcomp}
\usepackage{afterpage}
\usepackage[colorlinks=true, linkcolor=blue, citecolor=blue, filecolor=magenta]{hyperref}
\usepackage{url}
\usepackage{titlesec}
\usepackage{lipsum}

\begin{document}


\title{Experimental Side-Channel-Secure Quantum Key Distribution over 200 km}

\author{Yang Zhou}\thanks{These authors contributed equally to this work.}
\author{Jing-Yang Liu}\thanks{These authors contributed equally to this work.}
\author{Chun-Hui Zhang}\thanks{These authors contributed equally to this work.}
\author{Chun-Wang Yi}
\author{Teng-Wen Jiang}
\author{Le-Chen Xu}
\author{Qin Wang}\email{qinw@njupt.edu.cn}

\affiliation{Institute of Quantum Information and Technology, Nanjing University of Posts and Telecommunications, Nanjing 210003, China}
\affiliation{Key Lab of Broadband Wireless Communication and Sensor Network Technology, Ministry of Education, Nanjing University of Posts and Telecommunications, Nanjing 210003, China}



\begin{abstract}
Quantum key distribution (QKD) enables two remote parties to share encryption keys with information-theoretic security guaranteed by physical laws. Side-channel-secure QKD (SCS-QKD) has attracted considerable attention because it simultaneously removes source and detector side-channel vulnerabilities. Although a recent experiment demonstrated SCS-QKD over 50 km, practical implementation remains challenging due to imperfect vacuum preparation and finite-key constraints under coherent attacks. Here, following the theoretical framework of Jiang \emph{et al.} [Phys. Rev. Res. 6, 013266 (2024)], we experimentally implement a practical SCS-QKD protocol using an imperfect whole-space source and rigorous finite-key analysis. Benefiting from a stable GHz-level system operating at 1.25 GHz, we extend the transmission distance to 200 km and achieve high secure key rates of 18.31 kbps, 2.55 kbps, and 196.03 bps at 100 km, 150 km, and 200 km, respectively. Our results establish a new distance record for SCS-QKD and demonstrate the feasibility of high-speed, long-distance, and practically secure quantum key distribution.

\end{abstract}
\maketitle


\section{Introduction}
Quantum key distribution (QKD) enables two remote users (Alice and Bob) to share secret keys with information-theoretic security \cite{QKD1,QKD2,QKD3,QKD4,QKD6,QKD7}. While unconditional security has been rigorously established, practical implementations inevitably deviate from idealized assumptions, and device imperfections may introduce exploitable side channels \cite{QKD5}. Significant efforts have therefore been devoted to closing device loopholes \cite{effort1,effort2,effort3,effort4,effort5}. On the source side, the decoy-state method \cite{DECOY1,DECOY2,DECOY3} mitigates photon-number-splitting attacks \cite{PNS1,PNS2} and enables secure key generation with weak coherent sources. On the detection side, measurement-device-independent QKD (MDI-QKD) \cite{MDI1,MDI2,MDI3,MDI4,MDI5,MDI6,MDI7,MDI8,MDI9} removes all trust assumptions on measurement devices. Building upon the MDI architecture, twin-field QKD (TF-QKD) \cite{TF1,TF2,TF3,TF4,TF5,TF6,TF7,TF8,OPLL,TF9,TF10,TF11} further extends the achievable transmission distance and has been shown to overcome the Pirandola–Laurenza–Ottaviani–Banchi (PLOB) bound \cite{PLOB}.

Despite these advances, hidden side channels associated with imperfect state preparation remain a fundamental concern. In practice, the emitted photon pulses do not strictly reside in an ideal two-dimensional Hilbert space. Additional degrees of freedom—such as spectral components, temporal modes, or waveform distortions—may unintentionally encode information and be exploited by an eavesdropper \cite{QKD5,sidech1,sidech2,sidech3,sidech4,sidech5}. Conventional security analyses typically require detailed characterization of the emitted states, which is experimentally demanding and may be incomplete in high-dimensional physical spaces.

Side-channel-secure QKD (SCS-QKD) \cite{SCS0} was proposed to address this issue from a fundamentally different perspective. Instead of characterizing all physical degrees of freedom, the protocol establishes security by mapping imperfectly prepared practical states to an equivalent ideal source model. By incorporating the MDI architecture \cite{MDI1,MDI2}, SCS-QKD simultaneously eliminates detection-side vulnerabilities. An initial fiber demonstration over 50 km validated the feasibility of the scheme \cite{SCS50}. However, the original SCS protocol requires perfect vacuum states, which is unattainable in practice due to the finite extinction ratio of intensity modulators \cite{SCS50,SCS1}. To overcome this limitation, Jiang \emph{et al.} \cite{SCS2} proposed a practical SCS-QKD protocol that accommodates imperfect vacuum sources. Crucially, the security proof relies only on the experimentally measurable projection probability onto the vacuum state, rather than full knowledge of the emitted quantum states in the entire Hilbert space. This whole-space mapping approach preserves the security of the original SCS framework while significantly relaxing experimental requirements, making realistic implementations feasible.

Despite these theoretical advances, experimentally validating practical SCS-QKD over long distances remains challenging. Long fiber links demand stringent phase stabilization, accurate control of extinction ratios, and careful finite-size statistical analysis. Furthermore, achieving GHz repetition rates while maintaining security under realistic source imperfections has not yet been demonstrated over extended distances. In this work, we report a 1.25 GHz implementation of practical SCS-QKD over 200 km of optical fiber with real-time phase compensation. By employing optical injection locking (OIL) \cite{OIL1,OIL2} to distribute a common phase reference between Alice and Bob, stable phase coherence is achieved between remote users. Under finite-key analysis against coherent attacks, we obtain a secure key rate (SKR) of 196.03 bps at 200 km. Our results experimentally validate the practical whole-space mapping paradigm at high repetition rate and long distance, bridging the gap between theoretical source-side security and deployable quantum communication systems.

\section{Results}
\subsection{Principle of side-channel-secure QKD}
\begin{figure}[]
	\centering
	\includegraphics[width=1.0\linewidth]{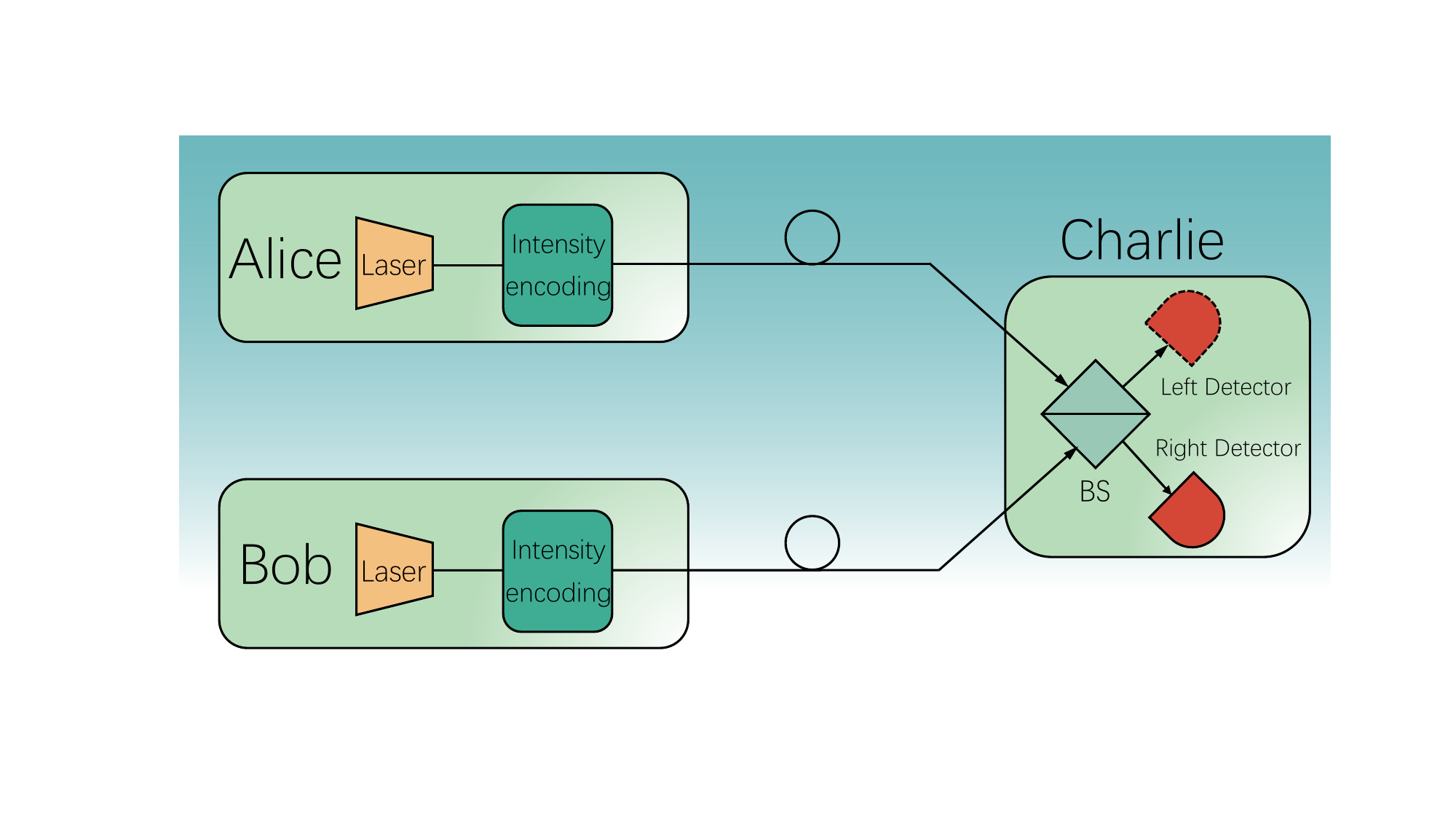}
	\caption{Schematic representation of the side-channel-secure quantum key distribution. Alice and Bob use intensity encoding to choose whether to send a coherent state pulse or a vacuum state pulse, while Charlie performs interference measurements  using a 50:50 beam splitter (BS) and two detectors.}
	\label{fig1}
\end{figure}

We adopt the practical SCS-QKD protocol \cite{SCS2} and the primary steps of the protocol demonstrated in this paper are summarized as follows. There are only two weak coherent state sources from Alice (Bob): the weak source $o_A$ ($o_B$) with intensity $\mu=\mu_{oA}\ (\mu_{oB})$, and the strong source $x_A$ ($x_B$) with intensity $\mu_A\ (\mu_B)$. For the time window $i$, Alice (Bob) randomly prepares a nonrandom phase coherent state from  sources $o_A$ or $x_A$ ($o_B$ or $x_B$) with probability $p_o$ and $p_x=1-p_o$, respectively. Alice (Bob) assigns a classical bit value of 1 (0) when choosing source $x_A$ \ ($x_B$), and a bit value of 0 (1) when selecting source $o_A$ \ ($o_B$) locally. Then Alice and Bob send their states to the middle node Charlie, who measures the interference of the coming states, as pictured in Fig. \ref{fig1}. In order to compensate for the fiber channel phase difference and thus obtain a small phase error rate, Alice and Bob are allowed to send strong phase reference pulses to Charlie at the same time. In reality, it is not possible to prepare ideal quantum states due to the presence of side channels in sources, but SCS-QKD protocol can ensure source security under real-world conditions with the following requirements \cite{SCS2}:
\begin{equation}
	\begin{aligned}
\langle 0| \rho_{o}|0\rangle \geq a_{o0} \geq 0.5, \quad\langle 0| \rho_{x}|0\rangle \geq a_{0} \geq 0.5, \\
\langle 0| \sigma_{o}|0\rangle \geq b_{o0} \geq 0.5, \quad\langle 0| \sigma_{x}|0\rangle \geq b_{0} \geq 0.5,
    \end{aligned}
\end{equation}
where $|0\rangle$ is the vacuum state in Fock space; $\rho_{o}$ or $\rho_{x}$ ($\sigma_{o}$ or $\sigma_{x}$) is the whole-space state of the pulse from source $o_A$ or $x_A$ ($o_B$ or $x_B$); $a_{o0}, a_0, b_{o0}, b_0$ are the known value to Alice and Bob.

Charlie will publicly announce the measurement results to Alice and Bob. Additionally, active phase compensation is implemented at the measurement station, ensuring interference between the two weak coherent states from Alice and Bob on the left single-photon detectors. If only the right detector clicks, Charlie declares a successful measurement, and Alice and Bob consider the $i$-th time window as effective. This occurrence is termed an effective event, and the corresponding bit is designated as an effective bit.

Alice and Bob repeat the above process $N$ times. In a given time window, if either Alice or Bob sends a pulse from strong sources, it is denoted as the $\mathcal{Z}$ window. If both send pulses from strong sources, it is denoted as the $\mathcal{B}$ window, and if both send pulses from weak sources, it is denoted as the $\mathcal{O}$ window. The bits in $\mathcal{Z}$ windows are classified as untagged bits.  

Finally, Alice and Bob can distill the secure keys according to the following formula by conducting error correction and privacy amplification \cite{SCS2,SCS3}:
\begin{align}\label{eq3}
	R =&\ \frac{1}{N}\left\{n_{\mathcal{Z}}[1-H(e_{ph})]-fM_{\mathcal{S}}H(E_{t})-\log_2\frac{2}{\varepsilon_{cor}} \right. \nonumber \\
	&- \left. 2\log_2\frac{1}{\varepsilon_{PA}}-7\sqrt{n_{\mathcal{Z}}\log_2\frac{2}{\bar{\varepsilon}}}\right\},
\end{align}
where $H(x)=-x\log_2x-(1-x)\log_2(1-x)$ is the binary Shannon entropy function; $n_{\mathcal{Z}}$ is the number of effective $\mathcal{Z}$ windows; $e_{ph}$ is the upper bound of the phase-flip error rate; $f$ is the error correction inefficiency which we set to $f = 1.1$; $M_\mathcal{S}$ is the number of survived bits and $E_{t}$ is the corresponding bit-flip error rate of the raw key strings; $\varepsilon_{cor}$ is the failure probability of the error correction; $\varepsilon_{PA}$ is the failure probability of the privacy amplification; $\bar{\varepsilon}$ is the coefficient of measuring the accuracy of estimating the smooth min-entropy.

To counter coherent attacks, the postselection strategy \cite{P} introduces a controlled shortening of the key length—specifically by $2(d^2-1)\log_2(N-1)$ bits—applied to keys initially distilled under collective attacks. By refining the collective attack-based key rate model through this adjustment, a unified formula addressing coherent attacks can be rigorously derived by
\begin{align}\label{eq4}
	R_{coh} =R-\frac{2(d^2-1)\log_2(N+1)}{N}
\end{align}
with the security coefficient $\varepsilon_{coh}=(\varepsilon_{cor}+\varepsilon_{PA}+\bar{\varepsilon}+3\epsilon)(N+1)^{d^2-1}$, which is set to $10^{-10}$ in the following text \cite{SCS2}. $d$ is the dimension of the local states shared by Alice and Bob, and $d$ = 8 in the SCS protocol. $\epsilon$ is the probability that the real value of a parameter lies outside of the chosen fluctuation range.

\subsection{Experimental implementation}
\begin{figure*}[]
	\centering
	\includegraphics[width=0.95\textwidth]{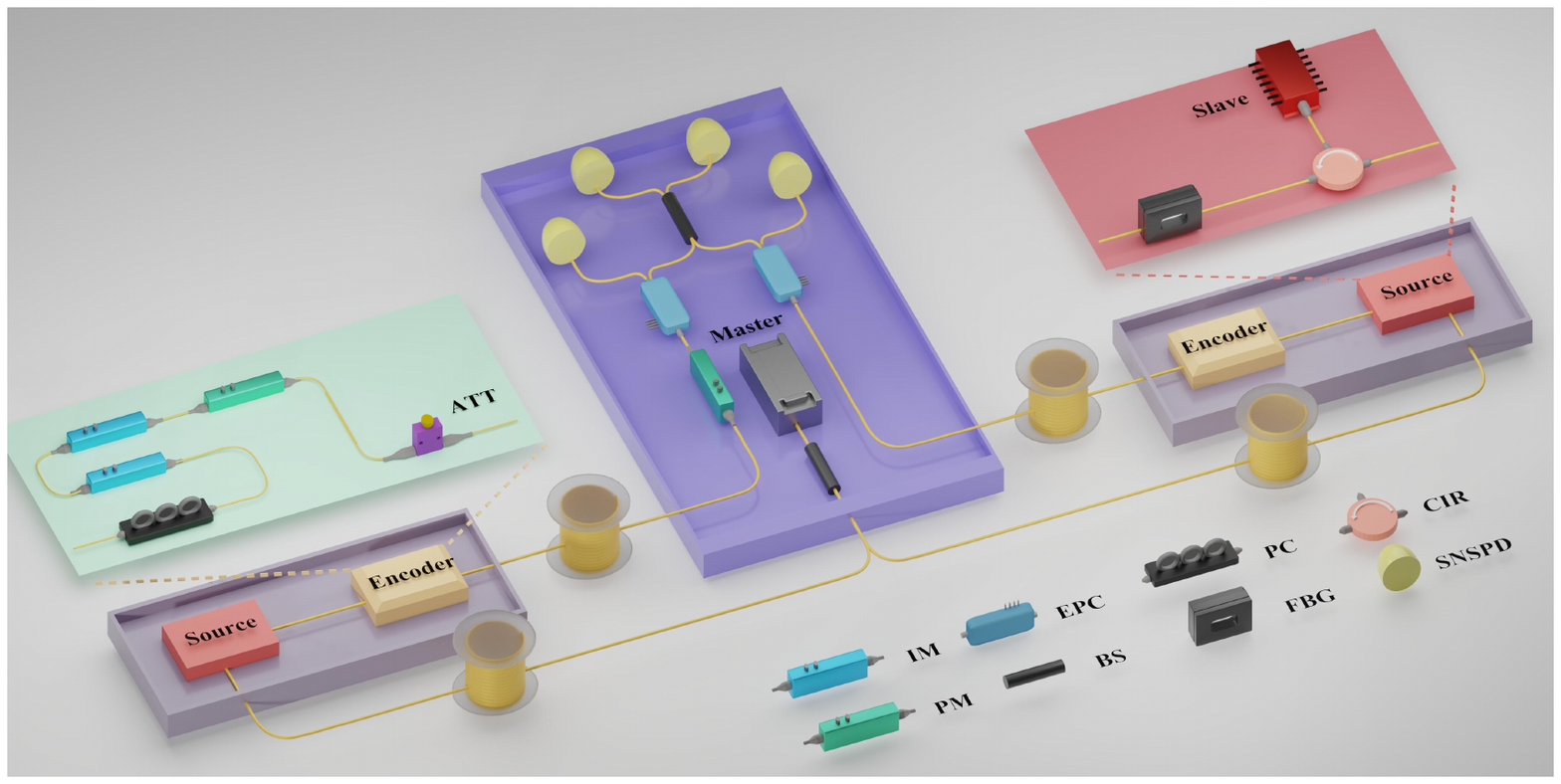}%
	\caption{Schematic of the experimental setup. A narrow-linewidth master laser at Charlie seeds injection-locked slave lasers at Alice and Bob, which are directly modulated to generate 1.25\,GHz optical pulses. The pulses pass through encoding modules composed of a polarization controller (PC), two intensity modulators (IMs) for coherent-state preparation and an attenuator (ATT), with a 10\,GHz fiber Bragg grating (FBG).  A phase modulator (PM) at Charlie implements active phase stabilization based on the two-phase scan scheme. Polarization is aligned by an electronic polarization controller (EPC), and the outputs are detected by superconducting nanowire single-photon detectors (SNSPDs). Circulator (CIR); Beam splitter (BS).}
	\label{fig2}
\end{figure*}

The experimental setup of SCS-QKD in our work is shown in Fig. \ref{fig2}, consisting of Alice, Bob, and the measurement node Charlie. Here, Alice and Bob each send out only one coherent state and one vacuum state, where the vacuum component is characterized through its projection probability, without assuming a perfectly ideal state.

The main experimental challenge is the stringent requirement for phase coherence between optical pulses emitted by distant and independent remote users. In practice, lasers located at separated stations operate independently, which makes it extremely difficult to maintain a stable and well-defined relative optical phase over long distances. To address this challenge, we adopt the OIL technique \cite{OIL1,OIL2} to distribute a common optical phase reference to the remote users. As shown in Fig. \ref{fig2}, a narrow-linewidth continuous-wave laser located at Charlie’s measurement station serves as the master laser. Its output is split by a 50:50 beam splitter (BS) into two seed beams and delivered to Alice’s and Bob’s stations through long optical fibers. At each station, a distributed-feedback (DFB) laser operates as a slave laser and is optically injected by the corresponding seed light. Under optical injection locking, the slave lasers are forced to emit optical fields that inherit the frequency and phase of the master laser, even when they are initially detuned by several times their intrinsic linewidth. As a result, the optical pulses generated at Alice’s and Bob’s sites originate from a common phase reference, thereby establishing phase coherence between distant users. 

In our experiment, the master laser (NKT Photonics Inc.) has a linewidth of approximately 0.1 kHz and emits continuous-wave light with a central wavelength of 1550.12 nm. The injection-locked slave lasers (Allwave Lasers Photonic Devices Inc.) are optically locked to the master laser. They are modulated by radio-frequency signals generated by an arbitrary waveform generator (AWG, Keysight 8190A) and amplified before driving the lasers, producing optical pulses at a clock rate of 1.25 GHz with a pulse width of approximately 390 ps. The master and slave lasers are connected via an optical circulator (CIR), which allows the seed light from Charlie to be injected into the slave lasers while simultaneously extracting the output pulses from the slave lasers for subsequent modulation and transmission. To further suppress unwanted spectral components, a 10 GHz fiber Bragg grating (FBG) is inserted into Alice’s and Bob’s apparatuses to filter the amplified spontaneous emission noise introduced during modulation and optical amplification. Once phase coherence between the remote lasers is established by optical injection locking, the optical pulses enter the encoding module. Two cascaded intensity modulators (IMs) are used to implement on–off modulation of the coherent states, determining whether a pulse is transmitted or suppressed. By jointly modulating the injection-locked pulses, the IMs prepare coherent states with an average photon number of 
$\mu_{A} \ (\mu_{B})$ for sending and vacuum states for not sending, while providing a high extinction ratio between the signal and vacuum states. The measured extinction ratios are 40.1 dB at 100 km, 35.7 dB at 150 km, and 30.5 dB at 200 km.

Another challenge is the accurate compensation of phase drift induced by long fiber channels. 
Here, active phase stabilization is implemented at Charlie using a phase modulator (PM). 
A frame period of $40\,\mu\mathrm{s}$ is employed, within which the optical pulses are time-multiplexed into a reference part and a quantum part. 
The reference window occupies $19.6\,\mu\mathrm{s}$ and consists of two equal intervals corresponding to two phase settings \{0,$\pi$/2\} that generate in-phase and quadrature interference, while the remaining $20\,\mu\mathrm{s}$ are used for quantum signal transmission. An additional $0.4\,\mu\mathrm{s}$ is reserved as a recovery interval for the feedback electronics. During the reference window, comparatively bright pulses are used to estimate the relative phase drift. 
The photon counts recorded by the two superconducting single-photon detectors (SNSPDs) under the two phase settings form a counting matrix, which is directly used to estimate the compensated phase voltage following the two-phase scan (2PS) scheme \cite{2PS}. The counts are accumulated by a $200\,\mathrm{MHz}$ field-programmable gate array (FPGA), which computes the required compensation value in real time. Finally, the pulses are attenuated to the desired levels via an electrical variable optical attenuator (ATT) and transmitted to Charlie through low-loss fibers, which have an average attenuation of less than 0.18 dB/km.

At Charlie’s station, a polarization compensation module aligns the polarization of the incoming pulses before single-photon interference at a 50:50 BS. The interference outcomes are detected by two SNSPDs and recorded by a time tagger. The SNSPDs used in this work exhibit a system detection efficiency of $69\%$ and dark counts below $0.1\,\mathrm{Hz}$.

\subsection{Performance over long distances}
\begin{figure}
	\centering
	\includegraphics[width=1.0\linewidth]{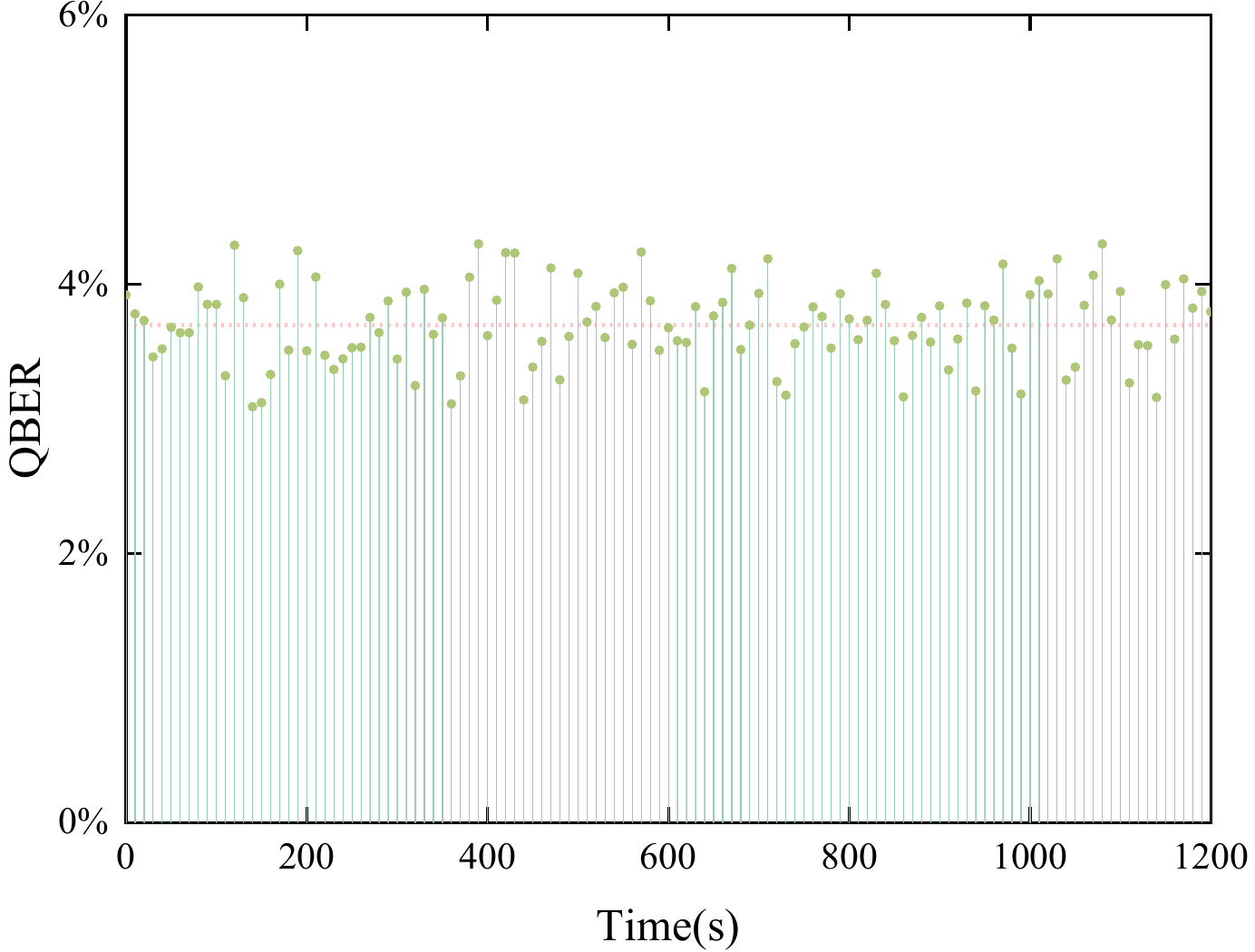}
	\caption{System stability over a 200 km fiber link. 
			The dots represent the measured quantum bit error rate (QBER) as a function of time when both Alice and Bob transmit coherent-state pulses under real-time phase compensation. 
			Each data point corresponds to the QBER accumulated over a 10 s time window. The dashed line indicates the average QBER of 3.6\%. }
	\label{figq}
\end{figure}

We demonstrate side-channel-secure QKD over 100 km, 150 km, and 200 km, with total losses of 18 dB, 27 dB, and 36 dB, respectively. Alice and Bob individually transmit $10^{12}$ pulses at 100 km, and $10^{13}$ pulses at both 150 km and 200 km. The parameters $\mu_A (\mu_B)$ and $p_x$ are globally optimized within the ranges [0, 0.5] and [0, 1], respectively, to maximize the secure key rate. In the finite-key regime, long-distance key generation requires stable system operation to accumulate sufficient quantum signals. In our experiment, stability is ensured by real-time active phase compensation. As shown in Fig.~\ref{figq}, the quantum bit error rate (QBER) over the 200 km fiber link remains stable over time when both Alice and Bob transmit coherent states, indicating reliable system performance under long-distance operation.

	\begin{figure}[!t]
	\centering
	\includegraphics[width=1.0\linewidth]{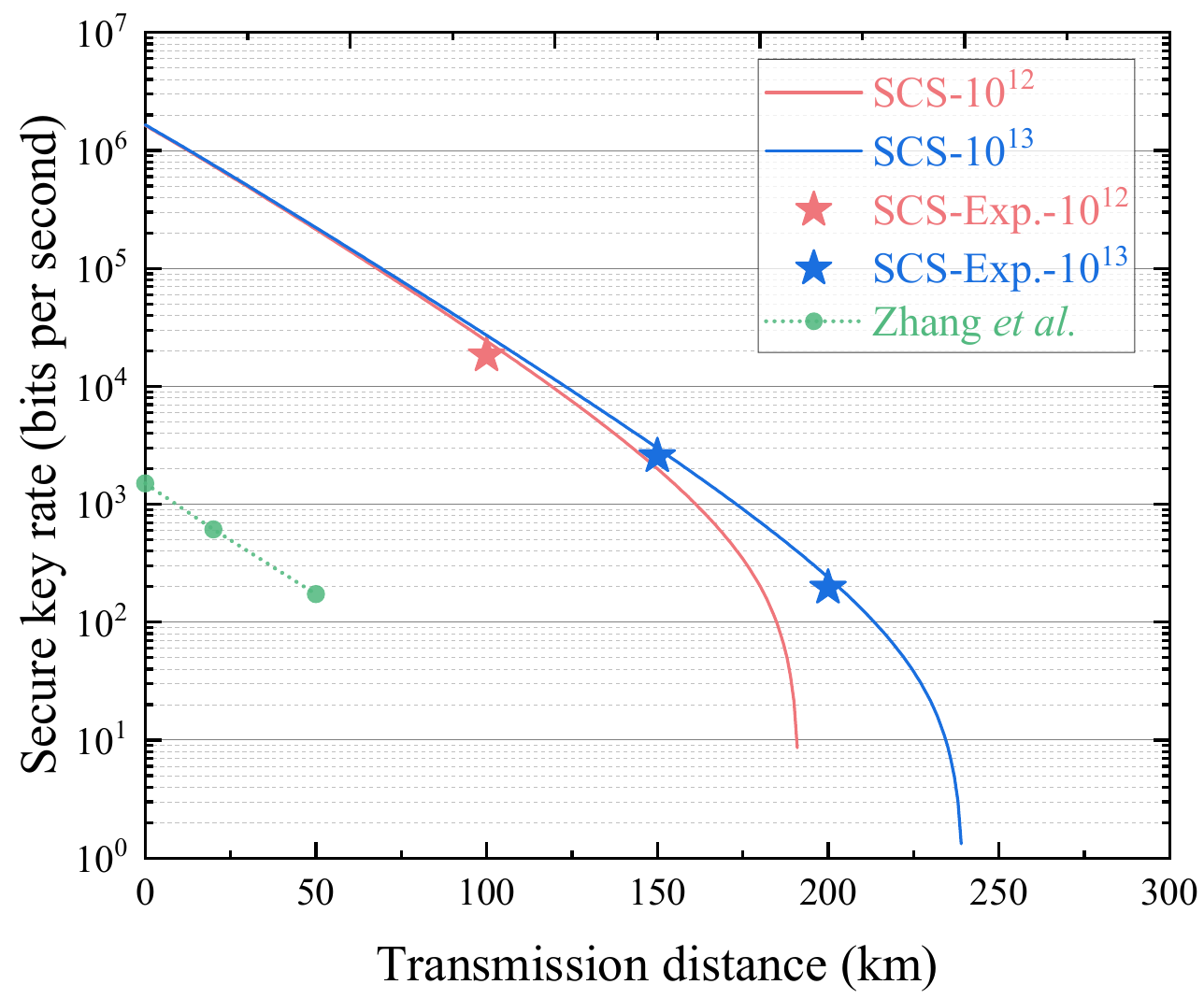}
	\caption{Secure key rate (bps) as a function of transmission distance between Alice and Bob. The red solid curve corresponds to the theoretical simulation of the practical SCS-QKD protocol with a total number of pulses $N=10^{12}$, while the blue solid curve represents the case of $N=10^{13}$. The corresponding star markers denote the experimental results. The simulations incorporate experimentally observed source intensity fluctuations of $\pm 0.65\%$. The green circle markers indicate previously reported state-of-the-art SCS-QKD experimental results from Ref.~\cite{SCS50}.
	}
	\label{fig5}
\end{figure}

We simulate the recently proposed practical scheme~\cite{SCS2} using optimized parameters. All simulations assume a uniform fiber attenuation of 0.18~dB/km, with full parameter optimization performed for each distance. The practical scheme is predicted to substantially enhance both the achievable key rate and the secure transmission distance. Figure~\ref{fig5} shows the experimental SKR as a function of fiber distance, together with theoretical simulations. At 150~km, the sifted key includes at least $2.29\times10^8$ untagged bits with a phase-flip error rate of 16.81\%. Under finite-key analysis, the secure key rate reaches 2.55 kbps. At 200~km, the sifted key includes at least $2.45\times10^7$ untagged bits with a phase-flip error rate of 16.72\%. Under finite-key analysis, the secure key rate reaches 196.03 bps. Overall, the experimental results agree well with theoretical predictions and demonstrate a clear performance enhancement across extended transmission distances.

In Ref.~\cite{SCS50}, the maximum demonstrated transmission distance was limited to 50~km, operating at a system repetition rate of 250~MHz, as indicated by the green circle in Fig. \ref{fig5}. Moreover, the security analysis required the preparation of an ideal vacuum state when Alice (Bob) chose not to send, which is practically unattainable due to the finite extinction ratio of intensity modulators. In contrast, our experiment implements the practical SCS-QKD protocol that rigorously maps imperfect source states to an equivalent ideal model, thereby removing the requirement of perfect vacuum preparation. Benefiting from this protocol framework and the enhanced system stability enabled by real-time phase compensation, together with a significantly increased repetition rate of 1.25~GHz, our system achieves substantially higher secure key rates and extends the transmission distance to 200~km. This demonstrates the feasibility of high-speed GHz-level SCS-QKD over long distances.

\section{Discussion}
We have experimentally demonstrated a practical SCS-QKD system under finite-key analysis, achieving secure key distribution over 200 km of optical fiber with a secret key rate of 196.03 bps. At a transmission distance of 100 km, the secret key rate reaches 18.31 kbps, representing an improvement of more than two orders of magnitude. Such a key rate is sufficient to support quantum-encrypted voice communication within metropolitan-area networks. Compared with previous implementations, our work significantly extends the transmission distance while operating at a GHz-level repetition rate, resulting in a substantial enhancement in overall system performance.

By integrating a practical security framework that rigorously incorporates source imperfections with a stable high-speed experimental platform, we verify that source-side side-channel security and measurement-device-independent security at the measurement site can be simultaneously realized between distant users. Importantly, this dual-end security paradigm is naturally compatible with network architectures in which untrusted nodes perform measurements while users retain trusted state preparation modules.

Our results demonstrate that practical source-side security based on the whole-space mapping framework can be implemented in high-speed and long-distance QKD systems, paving the way for scalable deployment in metropolitan and intercity quantum networks. This work provides an essential step toward realizing robust, high-rate, and side-channel-resilient quantum key distribution in future large-scale quantum communication infrastructures.

\section{Method}
\subsection{Theory of SCS-QKD protocol}
In carrying out the practical SCS-QKD protocol \cite{SCS2}, there are only two sources:  the weak source $o_A$ ($o_B$) with intensity $\mu=\mu_{oA} \ (\mu_{oB})$, and the strong source $x_A$ \ ($x_B$) with intensity $\mu_A \ (\mu_B)$. For the time window $i$, Alice (Bob) randomly prepares a nonrandom phase coherent state from  sources $o_A$ or $x_A$ ($o_B$ or $x_B$) with probability $p_o$ and $p_x=1-p_o$, respectively. In Ref. \cite{SCS2}, the theory has proven that the protocol of preparing the imperfect states \{$\rho_{o}, \rho_{x}$\} and \{$\sigma_{o}$, $\sigma_{x}$\} can be mapped by Eve from a protocol of preparing the perfect states \{$|0\rangle$, $|\sqrt{\mu_A}\rangle$\} and \{$|0\rangle$, $|\sqrt{\mu_B}\rangle$\}, where
\begin{equation}
	\begin{aligned}
		e^{-\mu_{A}}=\left|\sqrt{a_{0} a_{o0}}-\sqrt{\left(1-a_{0}\right)\left(1-a_{o0}\right)}\right|^{2}, \\
		e^{-\mu_{B}}=\left|\sqrt{b_{0} b_{o 0}}-\sqrt{\left(1-b_{0}\right)\left(1-b_{o0}\right)}\right|^{2},
	\end{aligned}
\end{equation}
and $|\sqrt{\mu_A}\rangle, |\sqrt{\mu_B}\rangle$ are coherent states with intensities $\mu_{A}, \mu_{B}$, respectively.
Next, we show how to calculate the key rate according to the real SCS-QKD protocol. 

We define the one-detector heralded events are called effective events. After Charlie announces those effective time windows to Alice and Bob, they respectively get $n_t$-bit strings, $Z_A$ and $Z_B$, formed by the corresponding bits of effective events of $\mathcal{Z}$ windows. 

In the ideal two-state protocol, the preparation in each time window can be equivalently described by the joint state
\begin{equation}
	\begin{aligned}
		|\phi\rangle = p_0 |00\rangle_S \otimes |01\rangle_I
		+ p_x |\mu_A \mu_B\rangle_S \otimes |10\rangle_I
		+ \sqrt{2 p_0 p_x}\, |\Psi\rangle,
	\end{aligned}
\end{equation}
where
\begin{equation}
	\begin{aligned}
		|\Psi\rangle = \frac{1}{\sqrt{2}}
		\left(
		|0 \mu_B\rangle_S \otimes |00\rangle_I
		+ |\mu_A 0\rangle_S \otimes |11\rangle_I
		\right).
	\end{aligned}
\end{equation}
Subsystem $I$ is kept locally by Alice and Bob, while subsystem $S$ is sent to Charlie. The procedure is repeated for $N$ rounds (subscripts $S$ and $I$ are omitted when no confusion arises). Charlie measures $|\phi\rangle^{\otimes N}$ and records whether each time window yields a click, storing the outcomes in a classical register $C$ accessible to Alice and Bob. Since SCS-QKD is measurement-device-independent, the measurement station is assumed to be fully controlled by Eve; thus Charlie is treated as Eve in the security analysis.

Under collective attacks, the shared state after $N$ rounds is $\rho_{ABC}^{\otimes N}$ with
\begin{equation}
	\begin{aligned}
		\rho_{ABC} = \mathrm{Tr}_{S E_{SC}}
		\left(
		|\phi\rangle\langle\phi|
		\otimes |{\rm click\ or\ not}\rangle\langle{\rm click\ or\ not}|_C
		\right),
	\end{aligned}
\end{equation}
where $E_{SC}(\cdot)$ denotes a map acting on subsystems $S$ and $C$. For each window, Alice and Bob first check $C$ to identify effective events, and then measure subsystem $I$ in the basis 
$\{|01\rangle\langle01|,\ |10\rangle\langle10|,\ |00\rangle\langle00|+|11\rangle\langle11|\}$ 
to determine the window type. The actual number of effective $\zeta$ windows is denoted by $\tilde n_\zeta$, where $\zeta\in\{O,B,Z\}$.

For effective $Z$ windows, subsystem $I$ is measured in the basis $\{|00\rangle,|11\rangle\}$ to extract the raw bit values of untagged bits. In the equivalent entanglement-based picture, the exact phase-flip error rate could in principle be obtained if subsystem $I$ were measured in the $X$ basis,
$\{|X^+\rangle=\frac{1}{\sqrt{2}}(|00\rangle+|11\rangle),\ |X^-\rangle=\frac{1}{\sqrt{2}}(|00\rangle-|11\rangle)\}$,
instead of the computational basis for effective $Z$ windows. A phase error occurs when Alice and Bob measure the state $|X^+\rangle$.

While the phase-flip error rate is not directly observable, it can be rigorously shown that, in the asymptotic limit, the expected number of phase errors satisfies\cite{SCS2}:
\begin{equation}
	\begin{aligned}
		\left\langle\bar{N}^{\mathrm{ph}}\right\rangle= & \frac{p_{0} p_{x}}{2}\left[\frac{c_{0}^{2}}{p_{0}^{2}}\left\langle n_{\mathcal{O}}\right\rangle^{U}+\frac{c_{1}^{2}}{p_{x}^{2}}\left\langle n_{\mathcal{B}}\right\rangle^{U}\right. \\
		& +\frac{2 c_{0} c_{1}}{p_{0} p_{x}} \sqrt{\left\langle n_{\mathcal{O}}\right\rangle^{U}\left\langle n_{\mathcal{B}}\right\rangle^{U}}+\frac{2 c_{0} \bar{c}_{2}}{p_{0}} \sqrt{N\left\langle n_{\mathcal{O}}\right\rangle^{U}} \\
		& \left.+\frac{2 c_{1} \bar{c}_{2}}{p_{x}} \sqrt{N\left\langle n_{\mathcal{B}}\right\rangle^{U}}+\bar{c}_{2}^{2} N\right],
	\end{aligned}
\end{equation}
where
\begin{equation}
	\begin{aligned} \bar{c}_{2}^{2}=(c_{0}+c_{1}-2 e^{-\mu_{A} / 2})(c_{0}+c_{1}-2 e^{-\mu_{B} / 2})
	\end{aligned}
\end{equation}
and we can take $c_0=e^{-(\mu_{A}+\mu_{B})/4}$, $c_1=e^{(\mu_{A}+\mu_{B})/4}$ \cite{SCS2}.

Then we can calculate the upper bound of the real value of the phase-flip error rate as
\begin{equation}
	\begin{aligned}
		{e}^{ph}&=\bar{N}^{ph}/n_{\mathcal{Z}},\\
		&=\phi^U(\left \langle \bar{N}^{ph}  \right \rangle)/n_{\mathcal{Z}},
	\end{aligned}
\end{equation}
and $\phi^U(x), \phi^L(x)$ are the upper and lower bounds while using Chernoff bound to estimate the real values according to the expected values. Let $X_1$, $X_2$,..., $X_n$ be $n$ random samples, detected with the value 1 or 0, and let $X$ denote their sum satisfying $X=\sum_{i=1}^{n}X_i$. $\phi$ is the expected value of $X$. We have \cite{chernoff1,chernoff2}
\begin{equation}
	\begin{aligned}
		\phi^{L}(X) & =\frac{X}{1+\delta_{1}(X)}, \\
		\phi^{U}(X) & =\frac{X}{1-\delta_{2}(X)},
	\end{aligned}
\end{equation}
where we can obtain the values of $\delta_1(X)$ and $\delta_{2}(X)$ by solving the following equations:
\begin{equation}
	\begin{aligned}
		\left[\frac{e^{\delta_{1}}}{\left(1+\delta_{1}\right)^{1+\delta_{1}}}\right]^{X /\left(1+\delta_{1}\right)}=\frac{\xi}{2}, \\
		\left[\frac{e^{-\delta_{2}}}{\left(1-\delta_{2}\right)^{1-\delta_{2}}}\right]^{X /\left(1-\delta_{2}\right)}=\frac{\xi}{2},
	\end{aligned}
\end{equation}
where $\xi$ is the failure probability.

In addition, we simulate the experimental observed values with the linear model presented in \cite{chernoff1} and assume a symmetric case in the SCS-QKD. If the total transmittance of the experiment setup is $\eta$, then we have
\begin{equation}
	\begin{aligned}
		n_{\mathcal{Z}}&=2p_0p_x(1-P_{dc})e^{-\eta\mu/2}[1-(1-P_{dc})e^{-\eta\mu/2}]N_{tot},\\
		n_{\mathcal{O}}&=p_0^2P_{dc}(1-P_{dc})N_{tot},\\
		n_{\mathcal{B}}&=p_x^2\{e_d(1-P_{dc})[1-(1-P_{dc})e^{-2\eta\mu}]  \\
		&+(1-e_d)P_{dc}(1-P_{dc})e^{-2\eta\mu}\}N_{tot},
	\end{aligned}
\end{equation}
where $P_{dc}$ is the dark count rate and $e_d$ is the optical misalignment error.

To better clarify the theoretical basis of the SCS-QKD protocol used in this work, we provide a detailed comparison between two versions of the SCS-QKD theory, as presented in Refs. \cite{SCS0} and \cite{SCS2}, respectively.

In the original SCS-QKD protocol \cite{SCS0}, the security is analyzed under the assumption that the quantum source emits pure states potentially containing side-channel information, modeled in an extended Hilbert space. It introduces a virtual–real source mapping framework, and Theorem~1 states that a real-life source, $\mathcal{S}$, emitting imperfect states in the whole space is equivalent to a perfect (virtual) source, $\mathcal{P}$, emitting perfect states, all of which are identical in the side-channel space if there exists a quantum process, $\mathcal{M}$, that can map source $\mathcal{P}$ to source $\mathcal{S}$. The final key of a QKD protocol using source $\mathcal{S}$ can be calculated by assuming that the virtual source, $\mathcal{P}$, is used. This result allows the security analysis to proceed as if ideal coherent states were sent, provided such a mapping exists. However, the existence of the mapping is not reduced to a clear quantitative condition; verifying it requires detailed knowledge of the global side-channel states. 

In contrast, the practical SCS-QKD protocol \cite{SCS2} significantly strengthens and generalizes the mapping-based security framework. A key theoretical advancement lies in providing a necessary and sufficient condition for mapping between mixed states: the protocol with actual source $\{\rho_1, \rho_2\}$ is secure, provided that the Bures fidelity $F(\rho_1, \rho_2) \ge |\langle \alpha_1 | \alpha_2 \rangle|$ for all time windows. The Bures fidelity $F(\rho_1, \rho_2)$ quantifies the closeness between two quantum states $\rho_1$ and $\rho_2$, and is defined as $F(\rho_1, \rho_2) =\mathrm{Tr} \left( \sqrt{ \sqrt{\rho_1} \rho_2 \sqrt{\rho_1} } \right).$ The states $|\alpha_1\rangle$ and $|\alpha_2\rangle$ refer to the \emph{ideal coherent states} that would be prepared in the corresponding time windows in a source without imperfections. Furthermore, Ref. \cite{SCS2} points out that the fidelity bound can be efficiently verified using only vacuum projection probabilities, noting that we only need the projection probability to vacuum for each real state rather than the exact information on what the real state is in the whole space. Based on this condition, Ref. \cite{SCS2} establishes a composable security proof in the finite-size regime, applying the postselection technique. The key rate with a finite number of pulses can be calculated because the postselection technique can be applied to the mapped perfect protocol. Moreover, the imperfect states for the real protocol here are not limited to pure states, i.e., they can be mixed states.

In summary, the original SCS-QKD protocol \cite{SCS0}, while providing a foundational theoretical framework for modeling side-channel imperfections via virtual–real source mappings, relied on pure-state assumptions and did not offer practical implementation criteria. In contrast, the practical SCS-QKD protocol \cite{SCS2} establishes a more general and experimentally feasible framework based on Bures fidelity, mixed-state modeling, and finite-key composable security analysis. 

Importantly, the difference in the key rate calculation between the two protocols arises primarily from the estimation of the phase-flip error rate $e_{\mathrm{ph}}$. Owing to their distinct conceptual constructions, the two protocols adopt different virtual entanglement representations, which lead to different bounds on $e_{\mathrm{ph}}$. A detailed derivation of the phase-flip error estimation can be found in Ref.~\cite{SCS2}.

Our work adopts this improved theoretical model, which is crucial in explaining the significantly higher secure key rates demonstrated in our experiment.

\subsection{Real-time phase compensation}

In long-distance fiber-based SCS-QKD, the relative optical phase between Alice and Bob undergoes slow fluctuations due to environmental perturbations such as temperature variations and mechanical vibrations. To maintain stable single-photon interference at Charlie, an active phase compensation scheme is implemented by applying feedback control to a phase modulator (PM).

The interference intensity at the beam splitter (BS) can be written as
\begin{equation}
	I=\frac{I_0}{2}\left[1+\cos(\phi)\right],
\end{equation}
where $I_0$ denotes the total intensity and $\phi$ represents the phase drift introduced by the fiber channel. Connecting one output port of the BS to detector $D_0$, the recorded photon counts can be written as
\begin{equation}
	C=C_0\cos(\phi)+C_0+C_d,
	\label{s9}
\end{equation}
where $C_0$ denotes the interference amplitude and $C_d$ represents the background contribution arising from detector dark counts and other imperfections. For the complementary output port connected to $D_1$, an additional $\pi$ phase shift leads to
\begin{equation}
	C=C_0\cos(\phi+\pi)+C_0+C_d.
	\label{s10}
\end{equation}
Within a short acquisition window, the background term can be neglected, allowing the relative phase offset to be inferred directly from detector statistics.

To determine the zero-phase voltage $V_0$ corresponding to constructive interference, a two-phase scan (2PS) procedure is performed. An initial probing voltage $V_i$ is applied to the PM, followed by a second voltage step $V_i+\frac{V_{\pi}}{2}$,
where $V_{\pi}$ denotes the half-wave voltage of the modulator. The photon counts recorded during the first and second voltage steps are denoted as $(N_0,M_0)$ and $(N_1,M_1)$, respectively, forming a counting matrix.

\begin{figure*}
	\centering
	\includegraphics[width=0.8\linewidth]{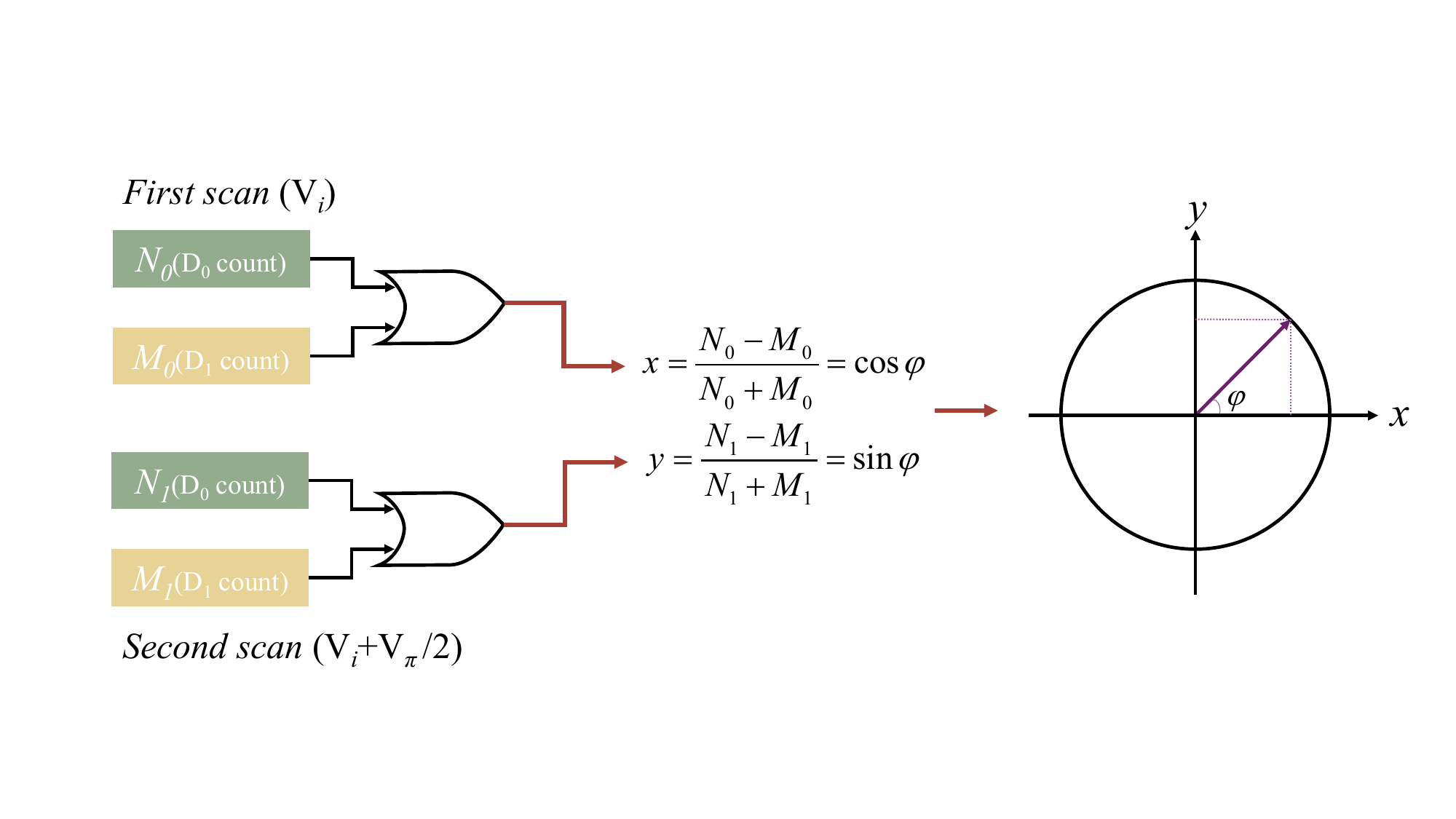}
	\caption{Schematic of the two-phase-scan (2PS) phase stabilization scheme. 
		In each reference frame, the FPGA sequentially records the photon counts from the two detectors under two orthogonal phase settings. 
		From these measurements, the normalized count differences are calculated to estimate the relative phase deviation between Alice and Bob. 
		Based on the reconstructed phase offset, the FPGA determines the required compensation value and outputs the corresponding control voltage to the phase modulator (PM), thereby completing the real-time feedback loop.
	}
	\label{fig4}
\end{figure*}

As illustrated in Fig. \ref{fig4}, based on  Eqs.~(\ref{s9}) and (\ref{s10}) we obtain
\begin{equation}
	\frac{N_0-M_0}{N_0+M_0}=\cos\left(\frac{V_c}{V_{\pi}}\pi\right),
\end{equation}
\begin{equation}
	\frac{N_1-M_1}{N_1+M_1}=\sin\left(\frac{V_s}{V_{\pi}}\pi\right).
\end{equation}
where $V_c$ and $V_s$ denote the voltage offsets from the zero-phase point obtained through $\arccos$ and $\arcsin$, respectively. Therefore, the measured pair $(x,y)$ forms a point on the unit circle, from which the phase offset can be reconstructed geometrically.

Accordingly, the candidate solutions for the compensated voltage are expressed as
\begin{equation}
	V'_0=
	\begin{cases}
		V_i - V_c, & \text{if } 1-\dfrac{2N_1}{N_1+M_1}>0,\\[6pt]
		V_i + V_c, & \text{otherwise},
	\end{cases}
\end{equation}

\begin{equation}
	V''_0=
	\begin{cases}
		V_i - V_s, & \text{if } \dfrac{2N_0}{N_0+M_0}-1>0,\\[6pt]
		V_i + V_s - V_\pi, & \text{otherwise}.
	\end{cases}
\end{equation}
In the absence of shot noise and systematic errors, the two estimates coincide. However, practical imperfections lead to small deviations between $V'_0$ and $V''_0$. Therefore, the final compensated voltage is taken as the average value
\begin{equation}
	V_0=\frac{V'_0+V''_0}{2}.
\end{equation}

The counts are accumulated by a $200\,\mathrm{MHz}$ field-programmable gate array (FPGA, Xilinx Zynq UltraScale+ MPSoCs EV), which computes the required compensation value in real time based on the above relations. The calculated control signal is sent to a 14-bit digital-to-analog converter (DAC, AD9767, Analog Devices) and subsequently amplified by a custom-built $15.75\times$ DC amplifier, providing a feedback voltage ranging from $-15.54\,\mathrm{V}$ to $+15.96\,\mathrm{V}$ to drive the PM. In addition, a 12-bit analog-to-digital converter (ADC, AD9238, Analog Devices) is used to acquire the detector signals for real-time processing.

After obtaining $V_0$, the feedback voltage is updated immediately to counteract the accumulated phase drift. By continuously repeating the 2PS and compensation procedure, the closed-loop system effectively suppresses channel-induced phase fluctuations and maintains high interference visibility during long-distance transmission.

\begin{figure*}
	\centering
	\includegraphics[width=0.9\linewidth]{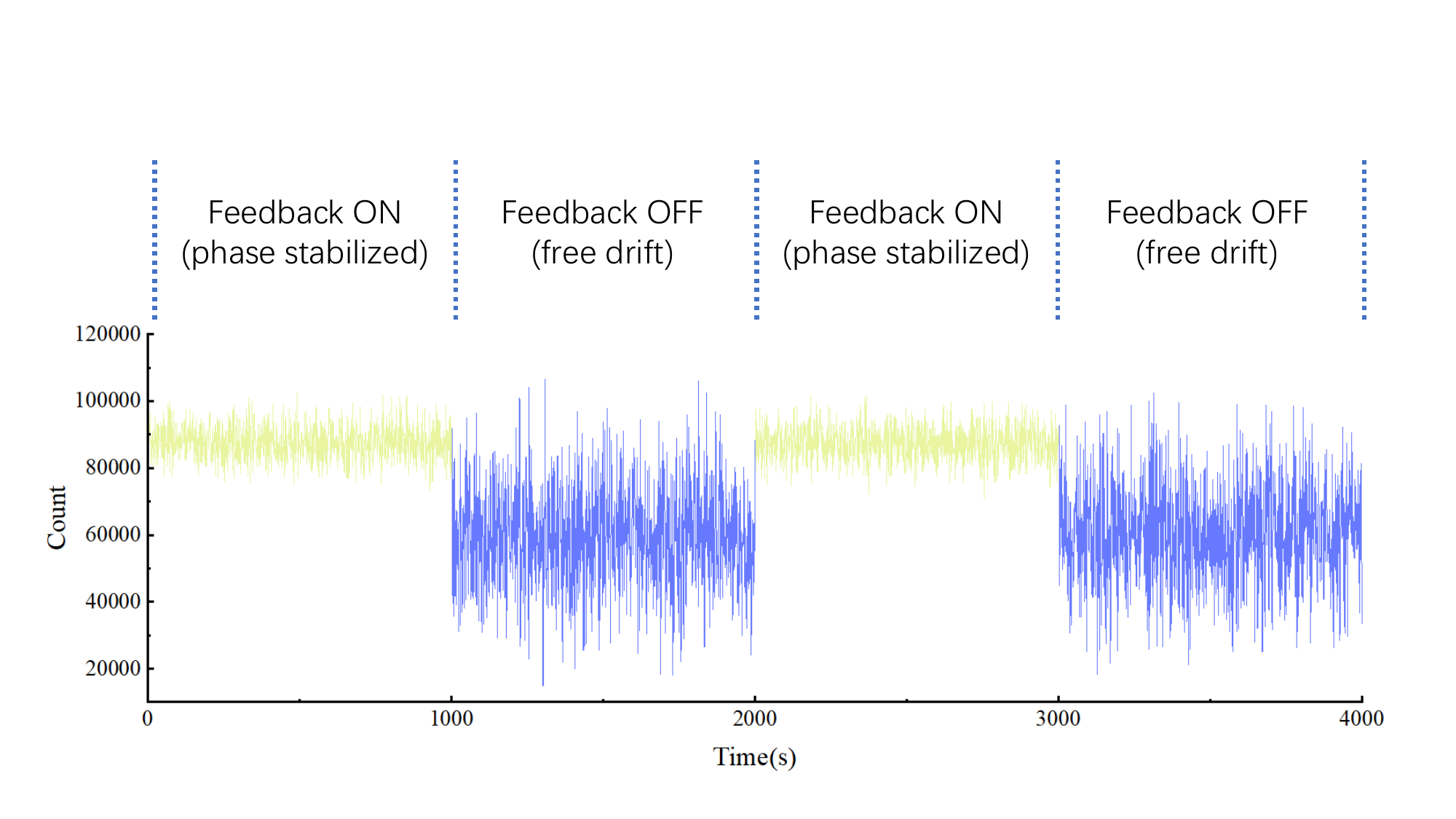}
	\caption{Photon counts from one detector with the phase feedback loop alternately enabled (ON) and disabled (OFF). With feedback ON, the counts remain stable around $90000$, indicating successful phase stabilization. When the feedback is OFF, the counts fluctuate significantly due to free phase drift in the fiber link.	}
	\label{fig6}
\end{figure*}

To evaluate the effectiveness of the real-time phase compensation, we experimentally compared the detector counts with the feedback enabled (ON) and disabled (OFF), as shown in Fig. \ref{fig6}. In this measurement, we continuously recorded the photon counts from one output port of the beam splitter. When the feedback loop is enabled, the detector counts remain stable around approximately $90000$ indicating that the relative phase is effectively locked near the constructive interference point. In contrast, when the feedback is turned off, the counts exhibit slow but significant fluctuations due to uncontrolled phase drift in the fiber channel. The free-running counts vary over a wide range, reflecting the gradual phase evolution induced by environmental perturbations such as temperature variations and mechanical vibrations. The phase estimation precision is limited by shot noise and finite counting statistics.

\emph{Acknowledgments.—} We gratefully appreciate Profs.
Xiang-bin Wang, Qiang Zhang, and Chuan-Feng Li for their enlighting discussions during this work. This work was supported by the Industrial Prospect and Key Core Technology Projects of Jiangsu Provincial Key R\&D Program (Grant No.~BE2022071), the Innovation Program for Quantum Science and Technology (Grant No.~2021ZD0300701), the National Natural Science Foundation of China (Grant Nos.~62471248, 62401287, 12074194), and the Natural Science Research Start-up Foundation for Introducing Talents of Nanjing University of Posts and Telecommunications (Grant No.~NY225014).

\emph{Author contributions.-}
Y.Z., J.-Y.L. and C.-H.Z. designed and performed the experiments. Y.Z. and J.-Y.L. carried out the theoretical analysis and numerical simulations. C.-W.Y., T.-W.J. and L.-C.X provided technical support and assisted with the experimental setup and data acquisition. Q.W. supervised the project. All authors contributed to data analysis, discussed the results, and approved the final version of the manuscript.

\emph{Data availability.-}
All data are available within the Article and Supplementary Files, or available
from the corresponding authors on reasonable request.

\emph{Competing interests.-}
The authors declare no competing interests.

\bibliography{SCS.bib}

\end{document}